# Generalizability issues with deep learning models in medicine and their potential solutions: illustrated with Cone-Beam Computed Tomography (CBCT) to Computed Tomography (CT) image conversion

Xiao Liang, Dan Nguyen, Steve Jiang*

Medical Artificial Intelligence and Automation Laboratory and Department of Radiation Oncology, University of Texas Southwestern Medical Center, Dallas, TX, USA

*E-mail: Steve.Jiang@UTSouthwestern.edu

**Abstract**

Generalizability is a concern when applying a deep learning (DL) model trained on one dataset to other datasets. It is challenging to demonstrate a DL model's generalizability efficiently and sufficiently before implementing the model in clinical practice. Training a universal model that works anywhere, anytime, for anybody is unrealistic. In this work, we demonstrate the generalizability problem, then explore potential solutions based on transfer learning by using the cone-beam computed tomography (CBCT) to computed tomography (CT) image conversion task as the testbed. Previous works only studied on one or two anatomical sites and used images from the same vendor's scanners. Here, we investigated how a model trained for one machine and one anatomical site works on other machines and other sites. We trained a model on CBCT images acquired from one vendor's scanners for head and neck cancer patients and applied it to images from another vendor's scanners and for prostate, pancreatic, and cervical cancer patients. We found that generalizability could be a significant problem for this particular application when applying a trained DL model to datasets from another vendor's scanners. We then explored three practical solutions based on transfer learning to solve this generalization problem: the target model, which is trained on a target domain from scratch; the combined model, which is trained on both source and target domain datasets from scratch; and the adapted model, which fine-tunes the trained source model to a target domain. We found that when there are sufficient data in the target domain, all three models can achieve good performance. When the target dataset is limited, the adapted model works the best, which indicates that using the fine-tuning strategy to adapt the trained model to an unseen target domain dataset is a viable and easy way to implement DL models in the clinic.

## 1. Introduction

Deep learning (DL) has been increasingly applied in medicine because it can improve the accuracy of diagnosis, prognosis, and treatment decision making by retrieving hidden information from big clinical data, improve efficiency by automating or augmenting clinical procedures, and transfer expertise to less experienced clinicians by learning from experienced clinicians. However, the generalizability of any given DL model must be demonstrated before that model can be implemented in clinical practice (Rajpurkar *et al.*, 2020). Many researchers train and test their DL models with their own data, but this gives no indication about the model's generalizability to other datasets. A better practice is to test the model with an external dataset, as some journals have recently started requiring for published DL research (David *et al.*, 2020). Although one external dataset is better than none, it still cannot sufficiently demonstrate the model's generalizability. Take, for example, a recent study on DL-based breast cancer screening (McKinney *et al.*, 2020). The investigators trained the model on two datasets from the UK and applied it to a single dataset from the US, then claimed that the DL system can generalize from the UK to the US. However, the US dataset came from only one institution, and 99% of the data were acquired from the same vendor's scanners. The test result for one vendor's scanners and one institution cannot represent all the clinical settings and data distributions for different institutions, vendors' scanners, scanning protocols, and so on. To address the problem of model generalizability, many researchers try to collect as much and as diverse patient data as possible to train a DL model that works in any clinical scenario, anytime, anywhere, for anybody. This ambitious goal seems unrealistic, as it is very challenging, if not impossible, to collect patient data from enough medical institutions to represent all clinical scenarios.

In this paper, we first illustrate the problem of generalizing a DL model, then we explore a few practical solutions based on transfer learning. We also investigate how the dataset scale influences the performance of the proposed solutions. This study uses image conversion from cone-beam computed tomography (CBCT) to computed tomography (CT) as the testbed. In cancer radiation therapy, CBCT images acquired during the treatment course are commonly used for positioning patients, monitoring anatomical changes, segmenting organs, and calculating radiation doses. However, CBCT's image quality is far inferior to CT's, mainly because of scatter and other artifacts. To improve CBCT's image quality, we previously proposed a CycleGAN model to convert CBCT to synthetic CT (sCT) images for patients with head and neck (H&N) cancer (Liang *et al.*, 2019). Subsequently, Harms et al. and Kida et al. also obtained similar results when using CycleGAN for patients with brain, pelvis and prostate cancers (Harms *et al.*, 2019; Kida *et al.*, 2019). These studies all focus on only one or two disease sites, and the CBCT images used for training and testing came from the same vendor's scanners. However, CBCT images might be scanned with different protocols or different machines, thus resulting in variant data distributions of image datasets and showing different displays in the same Hounsfield unit (HU) window, as illustrated in Figures 1 and 2. Thus, it is unknown whether a DL model trained with one dataset will work on another dataset. It is also unclear whether a DL model trained for one disease site will work on another. In this work, we conduct an extensive set of experiments using our CycleGAN model to perform the CBCT-to-CT conversion task with 7 different datasets from different disease sites and vendors' scanners. We first demonstrate the problem of model generalizability, taking our own model as exemplar, then we explore different methods to solve this problem.

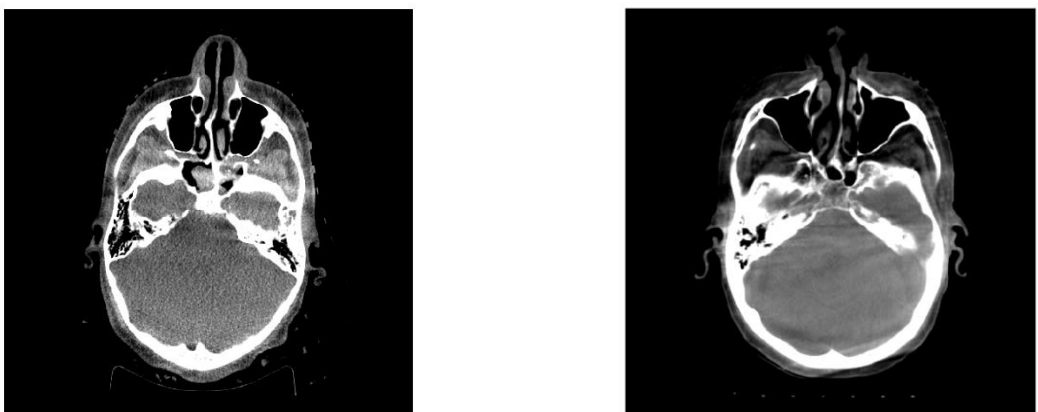

**Figure 1.** An illustration of different data distributions of CBCT datasets acquired with different scanning protocols from the same vendor's scanner. The CBCT images were acquired from a Varian OBI scanner with total exposure of 150 mAs (left) and 750 mAs (right). The left image is much noisier than the right one. The display window is from -200 HU to 400 HU.

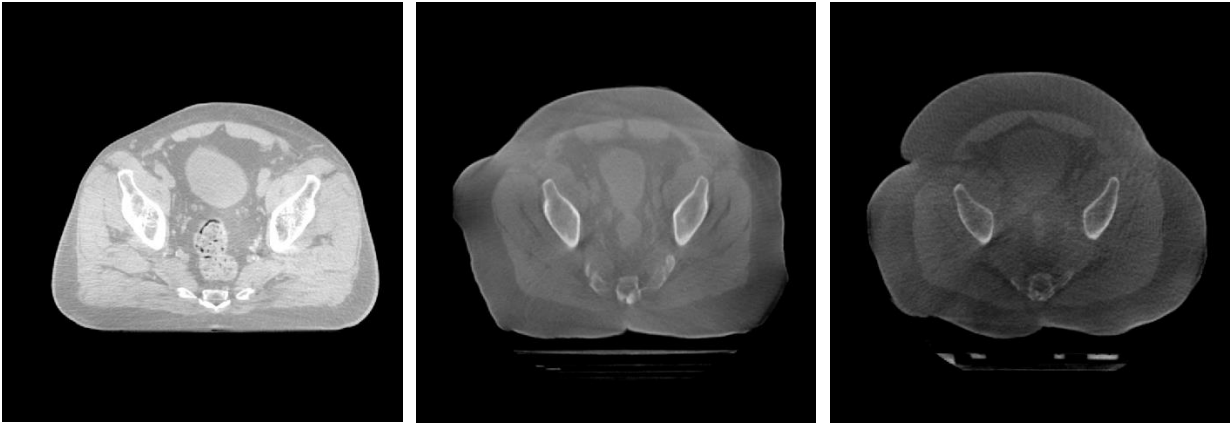

**Figure 2.** An illustration of different data distributions of CBCT datasets acquired from different vendors' scanners. The CBCT images were acquired with a Varian OBI scanner (left), Elekta XVI (Versa) scanner (middle), and Elekta XVI (Agility) scanner (right). In our institution, the XVI scanners on Versa and Agility machines have different hardware and software configurations, so we treat them as different scanners in this work. The display window is from -600 HU to 200 HU.

## 2. Data

We collected 7 datasets at our institution: two for patients with head and neck cancer (H&N1 and H&N2), three for patients with prostate cancer (Prostate1, Prostate2, and Prostate3), one for patients with cervical cancer (Cervix), and one for patients with pancreatic cancer (Pancreas), as shown in Table 1. Each dataset includes both CT and CBCT scans. CT scans in all 7 datasets were acquired via Philips CT scanner with the same kVp. Information about the CBCT scanners, vendors, and scanning protocols is given in Table 1.

The H&N1, H&N2, Prostate1, Prostate2, Prostate3, Cervix and Pancreas datasets contain 115, 22, 29, 54, 27, 28 and 28 patients, respectively. Each patient has a CT volume and a CBCT volume. Each CBCT volume in the H&N1, H&N2, Prostate1, Prostate2, Prostate3, Cervix and Pancreas datasets has 80, 80, 80, 70, 69, 69 and 70 image slices, respectively, with unified dimensions of $512 \times 512$. For training and validation, each CT volume was resampled to its corresponding CBCT's voxel spacing and then cropped to the CBCT's dimension and number of slices. Because our proposed CycleGAN did not require paired CT and CBCT images for training (Liang *et al.*, 2019), there was no need for image registration between CT and CBCT images. However, we performed rigid registration and deformable registration between CT and CBCT images in the testing dataset through commercial software (Velocity Oncology Imaging Informatics System, Varian Medical Systems, Inc.). The resulting deformed CT (dCT) images are our ground truth in this study and serve as the basis for evaluating the quality of the synthetic CT (sCT) images generated from the CBCT images via the CycleGAN model. The numbers of patients and 2D CBCT/CT images in each dataset for training, validation and testing are shown in Table 1.

**Table 1.** CBCT image datasets used for experiments.

| Dataset | Vender | Scanner | Scanning Protocol (kVp/mAs) | No. of patients for training/validation/testing | No. of images for training/validation/testing |
|---|---|---|---|---|---|
| **H&N1** | Varian | OBI | 100/150 | 83/9/23 | 6640/720/1840 |
| **H&N2** | Varian | OBI | 125/750 | 11/1/10 | 880/80/800 |
| **Prostate1** | Varian | OBI | 125/1070 | 15/3/11 | 1200/240/880 |
| **Prostate2** | Elekta | XVI (Versa) | 120/1600 | 15/3/11 | 1050/210/770 |
| **Prostate3** | Elekta | XVI (Agility) | 120/1600 | 15/2/10 | 1035/138/690 |
| **Cervix** | Elekta | XVI (Agility) | 120/1600 | 15/3/10 | 1035/207/690 |

| Pancreas | Elekta | XVI (Versa) | 120/1600 | 15/3/10 | 1050/210/700 |

## 3. Methods

### *3.1. Generalizability of the DL model*

This study used the CycleGAN architecture we developed previously. CycleGAN is a 2D network that can convert CBCT images to sCT images with less noise and other artifacts than previous architectures developed for this application (Liang *et al.*, 2019). Since the data distribution could be different for different disease sites, scanning protocols, and vendors' scanners, it is unknown whether a model trained on one dataset works for another dataset. To investigate the generalizability of the CycleGAN model trained on different CBCT datasets, we split the 7 datasets into source domains and target domains to mimic a situation where CBCT scans come from different clinical environments. The source domain consisted of the H&N1 and H&N2 datasets, which represented data collected from one circumstance; we used this domain to train a source model. We then applied the trained source model to five target domains—the Prostate1, Prostate2, Prostate3, Cervix and Pancreas datasets—to demonstrate the problem of the trained model's generalizability. To train the source model, we randomly split the H&N1 dataset into 83 and 9 patients and the H&N2 dataset into 11 and 1 patients for training and validation, respectively. Thus, the source model included 94 H&N patients in total for training and 10 patients in total for validation. Then, we compared the performance of the trained source model in both source and target domains to illustrate the generalizability problem that one might encounter. We tested the source model on 23, 10, 11, 11, 10, 10, and 10 patients from the H&N1, H&N2, Prostate1, Prostate2, Prostate3, Cervix, and Pancreas datasets, respectively.

### *3.2. Potential solutions to the generalizability problem*

To solve the generalizability problem mentioned above, we investigated three potential solutions: target model, combined model, and adapted model. The target model is trained on a target dataset starting from scratch. The combined model is trained on the combined source and target datasets starting from scratch. The adapted model is based on transfer learning and fine-tunes the trained source model on a target dataset. The source model trained on the H&N1 and H&N2 datasets was fine-tuned on the Prostate1, Prostate2, Prostate3, Cervix, and Pancreas datasets separately to get an adapted model for each target dataset. No layers in the architecture were frozen, and all layers were updated in the fine-tuning process. This strategy is commonly used for adapting an old model to a new domain when the training data from the new domain is limited.

We randomly split each target dataset into 15 patients for training, 2 or 3 patients for validation and 10 or 11 patients for testing. The testing datasets were the same as those in Section 3.1. The target training dataset was added to the source training dataset to train the combined model. Thus, 109 patients in total were used to train the combined model in each target domain. The validation and testing datasets of the combined model were the same as for the target model and adapted model.

For all three solutions, all hyper-parameters—including layer architecture, batch size and number of training iterations—were kept the same, except the learning rate. We used the grid search method to find the optimal learning rate for each target dataset (Bengio, 2012; Goodfellow *et al.*, 2016; Reed and MarksII, 1999). The search range of the learning rate was from 9E-3 to 2E-6.

### *3.3. Dependence of model performance on target dataset size*

The performance of the target, combined, and adapted models might depend on the size of the training dataset in the target domain. We conducted an experiment using Prostate2 dataset to investigate this effect. We trained the target, combined, and adapted models by randomly picking 5, 10, 15, 27 and 39 patients from the Prostate2 training dataset.

In the target and adapted models, the total number of patients for training was 5, 10, 15, 27 and 39. In the combined model, because the target dataset is added to the source dataset, the total number of patients for training was 99, 104, 109, 121 and 133. Eleven patients were used to test all the models, the same as in Sections 3.1 and 3.2. We repeated the training of each model 10 times for statistical analysis. We then analyzed the model performance in correlation with training data size for each model.

### 3.4. Evaluation methods

In addition to visually evaluating the image quality, we also evaluated model performance by using mean absolute error (MAE), root mean square error (RMSE), structural similarity index (SSIM), and signal-to-noise ratio (SNR), the four most widely used metrics (Al-Obaidi, 2015; Wang *et al.*, 2004) for image similarity measures. Assume $sCT(x, y)$ to be the HU value of the sCT images generated from the CycleGAN models and $dCT(x, y)$ to be the HU value of the deformably registered reference CT images. All the images have the same size [512,512].

MAE measures the difference between two images, as follows:

$$MAE = \frac{1}{n_x n_y}\sum_0^{n_x-1}\sum_0^{n_y-1}|dCT(x,y) - sCT(x,y)|. \qquad (1)$$

RMSE also measures the average errors between two images, but it gives relatively high weight to large errors, as follows:

$$RMSE = \sqrt{\frac{1}{n_x n_y}\sum_0^{n_x-1}\sum_0^{n_y-1}(dCT(x,y) - sCT(x,y))^2}. \qquad (2)$$

SSIM measures the similarity between two images based on the human visual system,(Wang *et al.*, 2003) that is

$$SSIM = [(2\mu_1\mu_2 + c_1)(2\sigma_{12} + c_2)]/[(\mu_1^2 + \mu_2^2 + c_1)(\sigma_1^2 + \sigma_2^2 + c_2)], \qquad (3)$$

where $\mu_1$ is the average pixel value of the dCT image, $\mu_2$ is the average pixel value of the sCT image, $\sigma_1^2$ is the pixel variance of the dCT, $\sigma_2^2$ is the pixel variance of the sCT, $\sigma_{12}$ is the pixel covariance of the dCT and sCT, $c_1 = (k_1 L)^2$, $c_2 = (k_2 L)^2$, L is the dynamic range of the pixel values (L=4095 in our case), $k_1 = 0.01$ and $k_2 = 0.03$. SSIM ranges from 0 to 1, and higher values indicate greater similarity between two images.

SNR compares the level of a desired signal to the level of background noise and is defined as

$$SNR = 10\log_{10}[\sum_0^{n_x-1}\sum_0^{n_y-1}(dCT(x,y))^2 / \sum_0^{n_x-1}\sum_0^{n_y-1}(dCT(x,y) - sCT(x,y))^2]. \qquad (4)$$

MAE and RMSE are given in HU. SSIM is given as a percent. SNR is given in dB.

## 4. Results

### 4.1. Performance of the source, target, combined, and adapted models

The quantitative evaluation results of the source model trained on the source domain datasets are shown in Tables 2 and 3. The generated sCT images and their corresponding CBCT and dCT images from the H&N1 and H&N2 testing datasets are shown in Supplementary Figures 1 and 2 for visual evaluation. We can see that the trained CycleGAN model can generate sCT images that are very similar to the reference dCT images and much better than the CBCT images when the model is applied to the same dataset that it was trained on.

**Table 2.** Evaluation of the source model performance on a source dataset (H&N1 dataset).

| | MAE (HU) | RMSE (HU) | SSIM (%) | SNR (dB) |
| --- | --- | --- | --- | --- |

|  | MAE (HU) | RMSE (HU) | SSIM (%) | SNR (dB) |
|---|---|---|---|---|
| CBCT vs. dCT | 59.13±16.71 | 152.44±31.50 | 80.50±5.93 | 15.41±2.02 |
| sCT vs. dCT | 36.81±9.13 | 94.14±14.57 | 86.11±3.91 | 19.26±1.67 |

**Table 3.** Evaluation of the source model performance on a source dataset (H&N2 dataset).

|  | MAE (HU) | RMSE (HU) | SSIM (%) | SNR (dB) |
|---|---|---|---|---|
| CBCT vs. dCT | 47.60±13.41 | 123.45±21.04 | 83.25±5.33 | 17.68±1.61 |
| sCT vs. dCT | 31.75±7.27 | 83.42±13.25 | 86.92±3.90 | 20.50±1.43 |

We quantitatively evaluated the performance of the source, target, combined, and adapted models in terms of MAE, RMSE, SSIM and SNR methods for each of the following target datasets: Prostate1, Prostate2, Prostate3, Cervix, and Pancreas (Tables 4-8). Paired sample t-tests were conducted between the adapted model and the other three models to evaluate the statistical significance of the differences in model performance. The sCT images generated by the source, target, combined, and adapted models and their corresponding CBCT and dCT images for each target dataset are shown in Supplementary Figures 3-7 for visual evaluation.

For the Prostate1 dataset, which comes from Varian OBI scanners, like the source datasets (H&N1 and H&N2), Table 4 shows similar MAE, RMSE, SSIM and SNR scores for the target, combined, and adapted models; all are only slightly better than those for the source model. This indicates that the source model performed reasonably well on this target dataset, and the three updated models performed comparably. The differences between the source and the adapted models' MAE scores (p=0.031) and SSIM scores (p=0.009) are small but statistically significant, thus indicating that the adapted model outperformed the source model. These results accord with Figure 3, where the lines of the target, combined and adaptive models coincide with each other, but the source model's line differs slightly from the three updated models and shows highest MAE and lowest SSIM. Thus, when applying the trained model to datasets collected from different disease sites but the same vendor's scanners, the source model generates good quality sCT images from CBCT, and the three updated models slightly improve upon this performance.

**Table 4.** Performance evaluation of four models on a target dataset (Prostate1). MAE, RMSE, SSIM, and SNR of CBCT and sCT images were calculated using dCT images as reference. The bold values represent the best results among the models. Paired sample t-tests were performed between the adapted model and the other models. Bold P-values indicate that the performance difference between the adapted model and one of the three other models is statistically significant (α=0.05).

|  |  | MAE (HU) | RMSE (HU) | SSIM (%) | SNR (dB) |
|---|---|---|---|---|---|
|  | CBCT | 43.78±6.93 | 106.03±14.01 | 80.11±4.83 | 17.84±2.14 |
| **Source model** | sCT | 21.92±7.57 | **55.32±10.92** | 88.77±5.49 | **23.34±2.00** |
|  | P-value | **0.031** | 0.171 | **0.009** | 0.179 |
| **Target model** | sCT | 20.13±5.70 | 57.29±10.84 | 91.18±3.03 | 23.01±1.88 |
|  | P-value | 0.106 | 0.830 | **0.004** | 0.772 |
| **Combined model** | sCT | **19.29±5.26** | 56.50±11.39 | **91.88±2.63** | 23.14±1.99 |
|  | P-value | 0.267 | 0.430 | 0.188 | 0.430 |
| **Adapted model** | sCT | 19.67±5.77 | 57.13±10.85 | 91.63±2.94 | 23.04±1.95 |

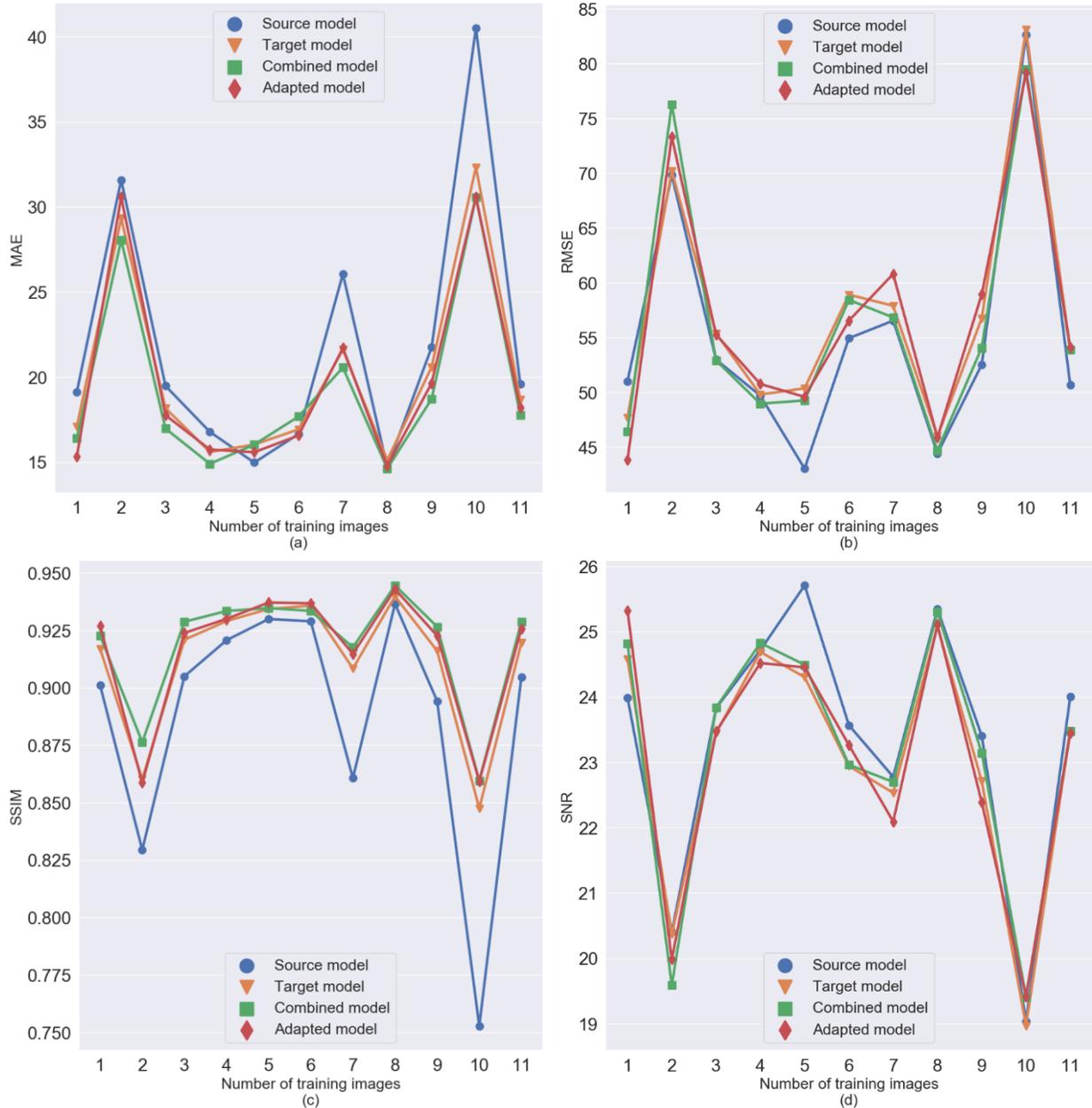

**Figure 3.** Comparison of the source, target, combined, and adapted models on the Prostate1 testing dataset: (a) MAE, (b) RMSE, (c) SSIM, and (d) SNR.

For the Prostate2, Prostate3, Cervix, and Pancreas datasets, which come from Elekta XVI scanners, the source model performed much worse in these target domains than in the Prostate1 target domain. The source model has the highest MAE and RMSE scores and the lowest SSIM and SNR scores among all the models (Tables 5-8). Upon visual evaluation of Figure 4, it is also clear that the lines of the source model are far away from those of the three updated models and show inferior performance in all metrics, which indicates that the source model performs poorly on the Prostate2 dataset. In contrast, the adapted model has the lowest MAE and RMSE scores and the highest SSIM and SNR scores among all the models (Tables 5-8). The paired sample t-tests comparing the MAE, RMSE, SSIM and SNR scores of the adapted model and the other three models all show a statistically significant difference. These results show that the adapted model has the best performance, which visual evaluation of Figure 4 also indicates. Thus, when applying the trained model to datasets collected from different disease sites and different vendors' scanners, the source model fails to generate visually good quality sCT images, and its MAE, RMSE, SSIM, SNR scores are substantially worse than in the previous scenario (Prostate1). All three updated models greatly outperform the source model, and the adapted model always performs best.

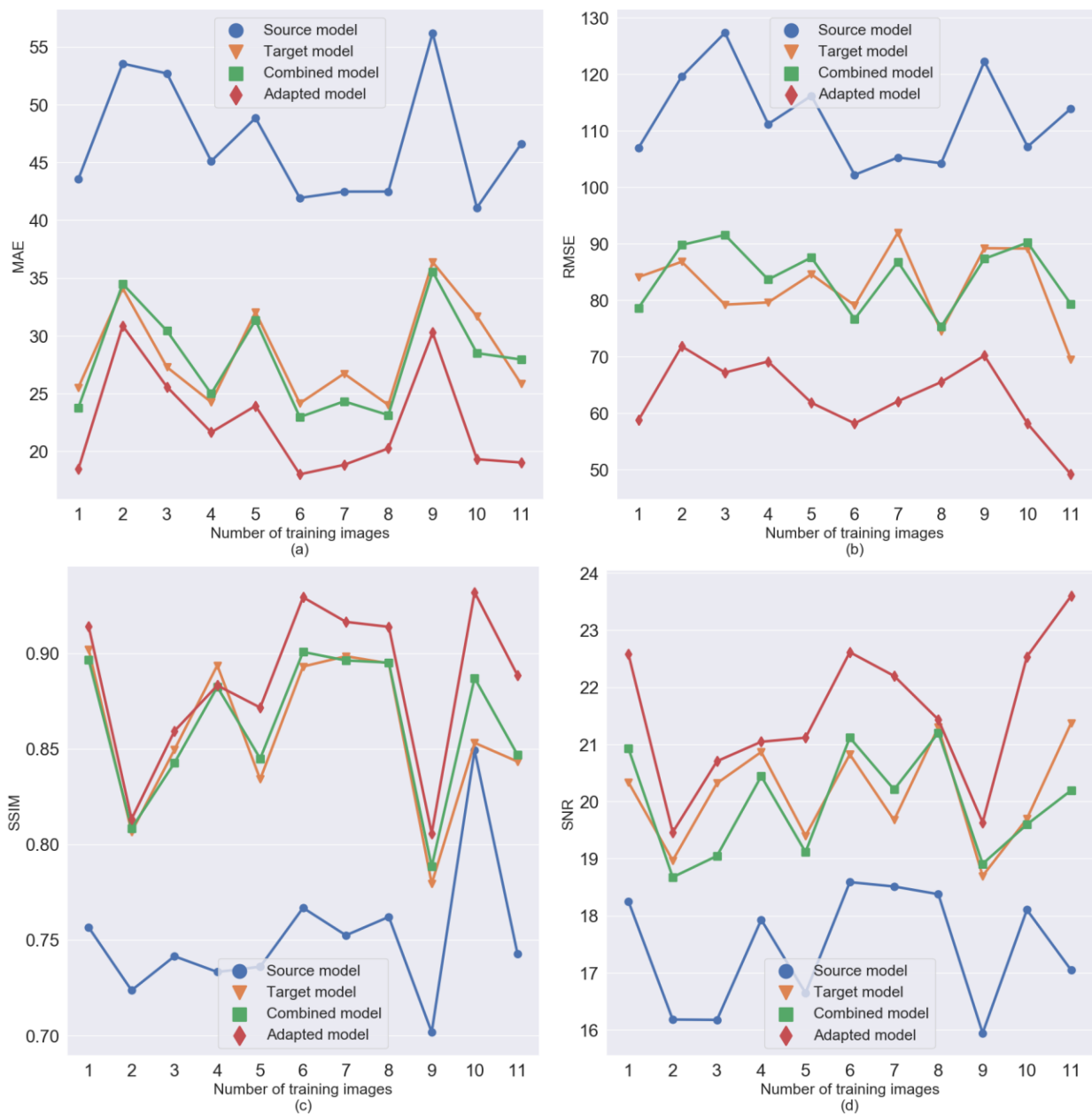

**Figure 4.** Comparison of the source, target, combined, and adapted models on the Prostate2 testing dataset: (a) MAE, (b) RMSE, (c) SSIM, and (d) SNR.

**Table 5.** Performance evaluation of four models on a target dataset (Prostate2). MAE, RMSE, SSIM, and SNR of CBCT and sCT images were calculated using dCT images as reference. The bold values represent the best results among the models. Paired sample t-tests were performed between the adapted model and the other models. Bold P-values indicate that the performance difference between the adapted model and one of the three other models is statistically significant ($\alpha=0.05$).

|  |  | MAE (HU) | RMSE (HU) | SSIM (%) | SNR (dB) |
| --- | --- | --- | --- | --- | --- |
|  | CBCT | 81.77±24.71 | 161.20±29.30 | 72.31±5.023 | 14.42±2.09 |
| Source model | sCT | 46.78±5.30 | 112.39±8.18 | 75.14±3.73 | 17.43±1.04 |
|  | P-value | **1.36E-12** | **1.46E-9** | **3.66E-8** | **7.17E-8** |
| Target model | sCT | 28.36±4.40 | 82.52±6.84 | 85.89±4.14 | 20.13±0.91 |
|  | P-value | **3.00E-4** | **7.85E-8** | **5.39E-3** | **9.34E-4** |
| Combined model | sCT | 27.95±4.54 | 84.26±5.83 | 86.27±3.90 | 19.95±0.93 |
|  | P-value | **4.04E-5** | **3.58E-11** | **3.26E-4** | **3.19E-4** |
| Adapted model | sCT | **22.38±4.69** | **62.93±6.70** | **88.43±4.37** | **21.54±1.30** |

**Table 6.** Performance evaluation of four models on a target dataset (Prostate3). MAE, RMSE, SSIM, and SNR of CBCT and sCT images were calculated using dCT images as reference. The bold values represent the best results among the models. Paired sample t-tests were performed between the adapted model and the other models. Bold P-values indicate that the performance difference between the adapted model and one of the three other models is statistically significant ($\alpha$=0.05).

|  |  | MAE (HU) | RMSE (HU) | SSIM (%) | SNR (dB) |
| --- | --- | --- | --- | --- | --- |
|  | CBCT | 119.89±34.89 | 220.91±38.55 | 70.95±6.88 | 11.61±2.08 |
| Source model | sCT | 60.49±10.38 | 132.43±12.90 | 74.47±5.49 | 15.96±1.37 |
|  | P-value | **3.03E-10** | **2.29E-8** | **2.76E-7** | **3.98E-7** |
| Target model | sCT | 31.51±8.14 | 100.85±16.45 | **84.58±4.05** | 18.41±1.85 |
|  | P-value | **7.40E-5** | **1.01E-4** | 0.083 | **1.95E-4** |
| Combined model | sCT | 33.17±9.15 | 92.85±12.82 | 82.40±5.86 | 19.09±1.62 |
|  | P-value | **2.04E-4** | **2.56E-4** | **5.84E-3** | **5.63E-4** |
| Adapted model | sCT | **26.95±8.46** | **76.89±15.37** | 85.26±4.68 | **20.83±2.16** |

**Table 7.** Performance evaluation of four models on a target dataset (Cervix). MAE, RMSE, SSIM, and SNR of CBCT and sCT images were calculated using dCT images as reference. The bold values represent the best results among the models. Paired sample t-tests were performed between the adapted model and the other models. Bold P-values indicate that the performance difference between the adapted model and one of the three other models is statistically significant ($\alpha$=0.05).

|  |  | MAE (HU) | RMSE (HU) | SSIM (%) | SNR (dB) |
| --- | --- | --- | --- | --- | --- |
|  | CBCT | 99.94±35.86 | 240.71±38.77 | 75.18±6.70 | 11.26±2.02 |
| Source model | sCT | 49.01±9.44 | 115.14±13.28 | 79.50±4.13 | 17.39±1.40 |
|  | P-value | **1.62E-8** | **5.00E-7** | **8.43E-8** | **2.70E-6** |
| Target model | sCT | 27.69±9.87 | 94.09±13.68 | 86.04±6.70 | 19.31±1.75 |
|  | P-value | **1.42E-4** | **1.37E-4** | **0.011** | **1.79E-4** |
| Combined model | sCT | 26.53±8.48 | 83.90±14.87 | 87.13±4.95 | 20.21±1.92 |
|  | P-value | **8.96E-5** | **5.29E-4** | **2.89E-3** | **7.49E-4** |
| Adapted model | sCT | **22.14±7.44** | **75.81±13.80** | **88.88±4.07** | **21.13±2.06** |

**Table 8.** Performance evaluation of four models on a target dataset (Pancreas). MAE, RMSE, SSIM, and SNR of CBCT and sCT images were calculated using dCT images as reference. The bold values represent the best results among the models. Paired sample t-tests were performed between the adapted model and the other models. Bold P-values indicate that the performance difference between the adapted model and one of the three other models is statistically significant ($\alpha$=0.05).

|  |  | MAE (HU) | RMSE (HU) | SSIM (%) | SNR (dB) |
| --- | --- | --- | --- | --- | --- |
|  | CBCT | 55.35±15.15 | 123.18±18.01 | 78.03±5.10 | 17.11±1.59 |
| Source model | sCT | 37.22±6.42 | 94.19±10.56 | 79.74±4.87 | 19.41±1.17 |
|  | P-value | **1.70E-8** | **1.27E-7** | **1.49E-7** | **2.70E-7** |
| Target model | sCT | 25.17±6.09 | 77.28±11.28 | 86.79±3.94 | 20.23±1.12 |
|  | P-value | **0.027** | **1.43E-3** | 0.663 | **2.36E-3** |
| **Combined model** | sCT | 26.31±4.93 | 85.644±8.03 | 86.69±3.81 | 20.23±1.05 |
| | P-value | 7.72E-3 | 5.48E-5 | 0.541 | 2.06E-4 |
| **Adapted model** | sCT | **23.56±4.95** | **75.78±9.76** | **87.39±3.81** | **21.34±1.35** |

### *4.2. Dependence of model performance on target dataset size*

Figure 5 shows that the performance of the target, combined, and adapted models depends on the size of the training dataset in the target domain (Prostate2). The three models performed similarly with a large number (2730) of training images. As the number of training images decreased, the accuracy of the combined model decreased the fastest, while the accuracy of the adapted model decreased the slowest. Because the combined model was trained on the Prostate2, H&N1 and H&N2 datasets, the training data for the combined model becomes unbalanced when there is much less Prostate2 data than H&N1 and H&N2 data. The knowledge previously learned from the H&N1 and H&N2 data gives the adapted model a good starting point to update features, so it requires less training data to achieve good performance. Therefore, with a small number of training images, the adapted model is clearly superior to the target and combined models. The adapted model still achieved good performance when trained on only 1050 images from 15 patients. These results suggest that model fine-tuning is the best option to guarantee good performance and robust training regardless of the amount of data available.

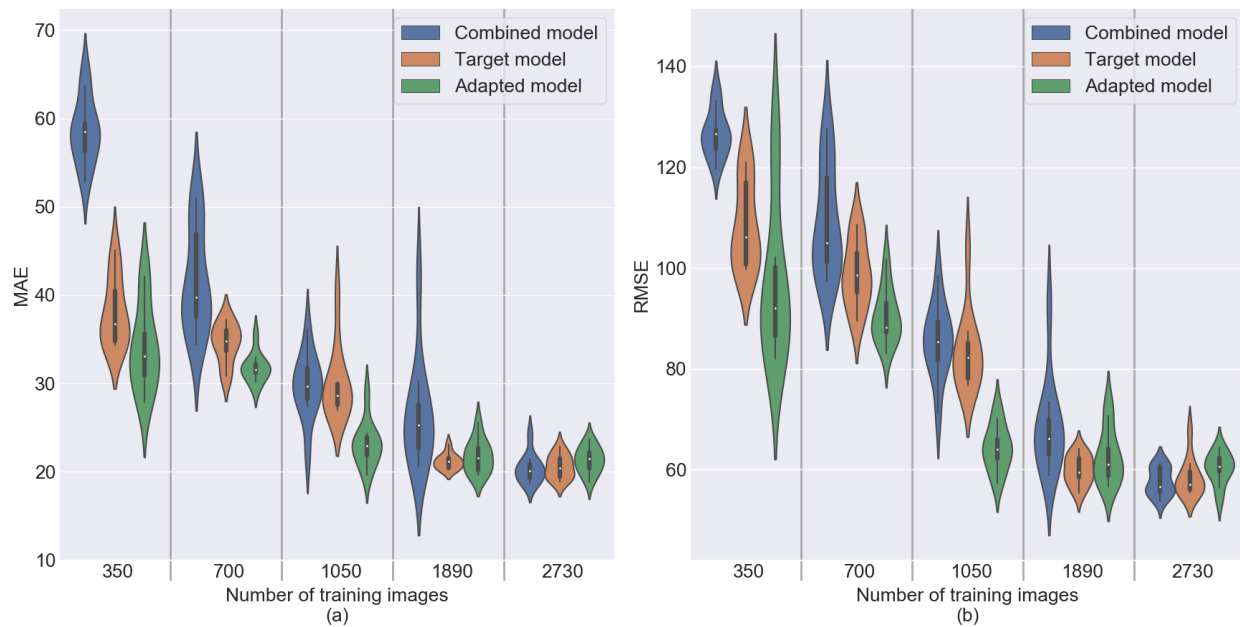

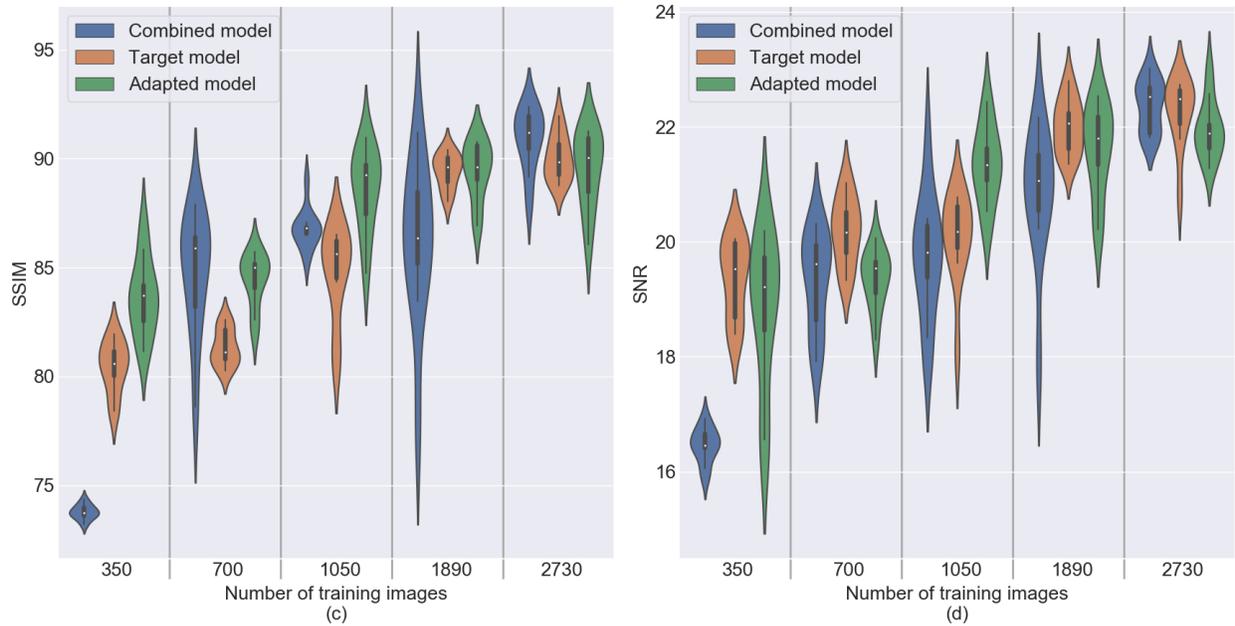

**Figure 5.** The influence of training data size on the performance of target, combined, and adapted models: (a) MAE, (b) RMSE, (c) SSIM, and (d) SNR.

## 5. Discussion and Conclusions

In this paper, we have illustrated the problem of generalizability for a DL model in a CBCT-to-CT image conversion application, and we showed that the model trained on the source domain does perform differently in different target domains. In our application, disease site was a minor influence on the source model's performance, but vendor's scanner was a major influence that could dramatically decrease the accuracy of the source model.

To solve this generalizability problem, we compared three solutions—target, combined and adapted models—in each target domain. The target model is trained on a target dataset starting from scratch. The combined model is trained on a combined source and target dataset starting from scratch. The adapted model fine-tunes the trained source model on the target dataset. We found that all three models modestly outperform the source model when the source model already performs well on a target dataset from the same vendor's scanners, but they significantly outperform the source model when the source model performs poorly on a target dataset from a different vendor's scanners. Among the three updated models, the adapted model works the best.

By analyzing the change in the models' performance with different numbers of training images, we found that the target, combined and adapted models perform comparably with a large number of training images, but the adapted model significantly outperforms the other two with smaller numbers of training images. Therefore, we suggest using the fine-tuning strategy to solve the generalization problem when deploying a DL model in clinical settings.

The generalizability of DL models is a challenging issue in medicine and is important for clinical implementation, where accuracy and precision are required. Even though this study only uses data from one institution, it has already shown a significant problem within one institution, so more significant problems should be expected for datasets from different institutions. Future studies should investigate this.

Another limitation of this work is that we only study one clinical task, CBCT-to-CT image conversion, to illustrate the problem and test three solutions. The number of training images needed to train or fine-tune a model is task-specific. However, we have shown the problem and solutions clearly, and we can still draw general conclusions from this one task.

We promote using the fine-tuning strategy for commissioning a model before implementing it in clinical practice, as it is very difficult to include enough data types from enough institutions to train a universal model. Future studies can focus on implementing the fine-tuning strategy through an automatic workflow that can be used easily in clinical environments by clinicians.

**Data availability**

All datasets were collected from one institution and are non-public. According to HIPAA policy, access to the dataset will be granted on a case by case basis upon submission of a request to the corresponding authors and the institution.

**Code availability**

The CycleGAN model algorithm is free to download for non-commercial research purposes on GitHub (https://github.com/lxaibl/CycleGAN-CBCT-to-CT).

**Acknowledgement**

We thank Dr. Jonathan Feinberg for editing the manuscript and Varian Medical Systems, Inc., for funding part of the project.

**Axial**                               **Sagittal**                                **Coronal**

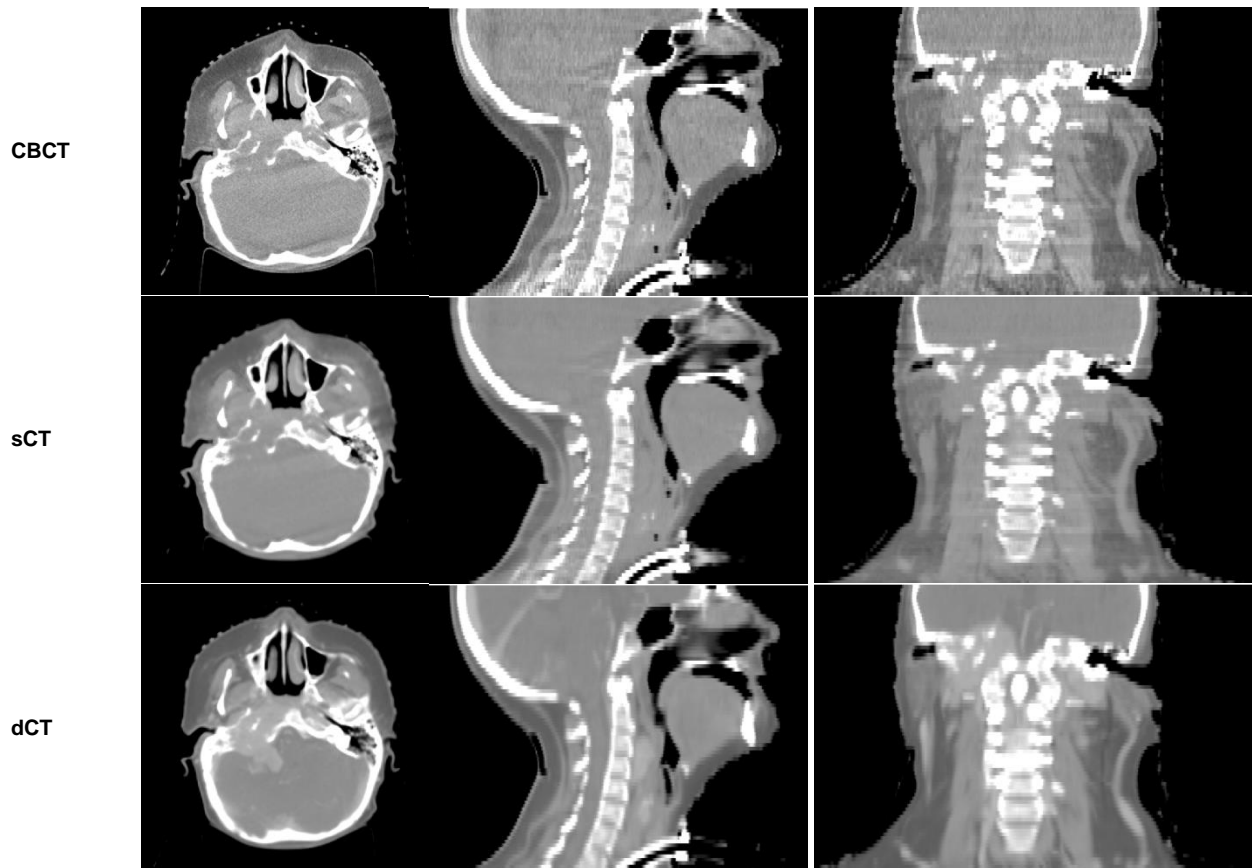

**Supplementary Figure 1.** Visualization of one randomly selected patient's CBCT, sCT and dCT images (H&N1). The top row shows CBCT images; the middle row shows sCT images generated from CBCT by the source model; and the bottom row shows the dCT images used as the ground truth. The left column shows images from the axial view; the middle column shows images from the sagittal view; and the right column shows images from the coronal view. The display window is from -400 HU to 400 HU.

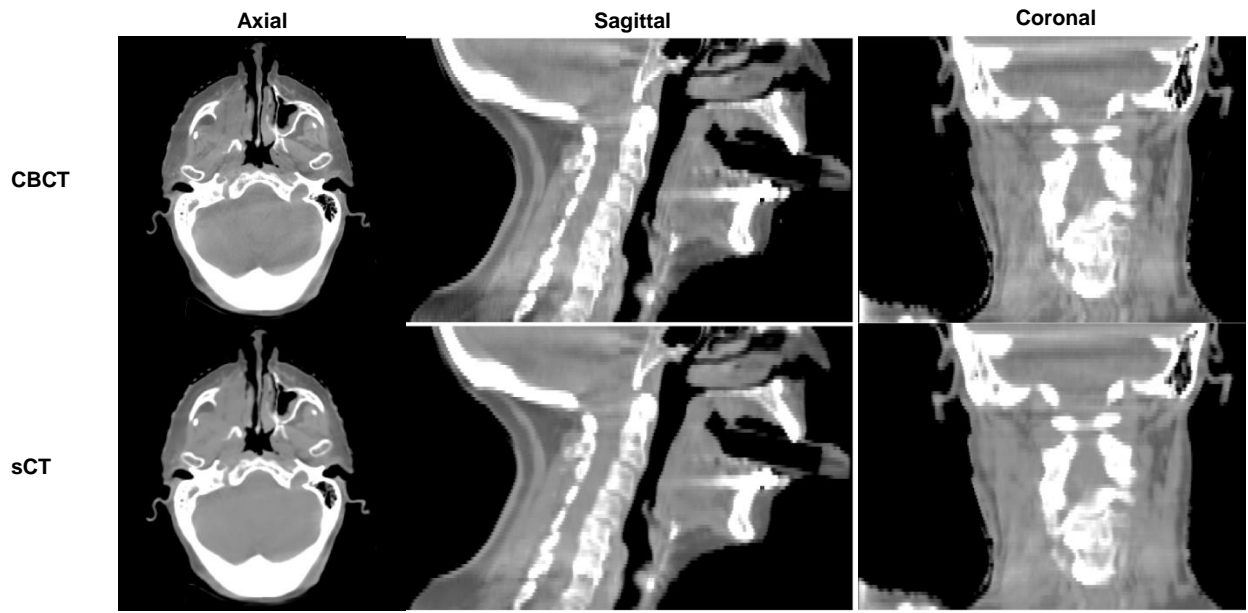

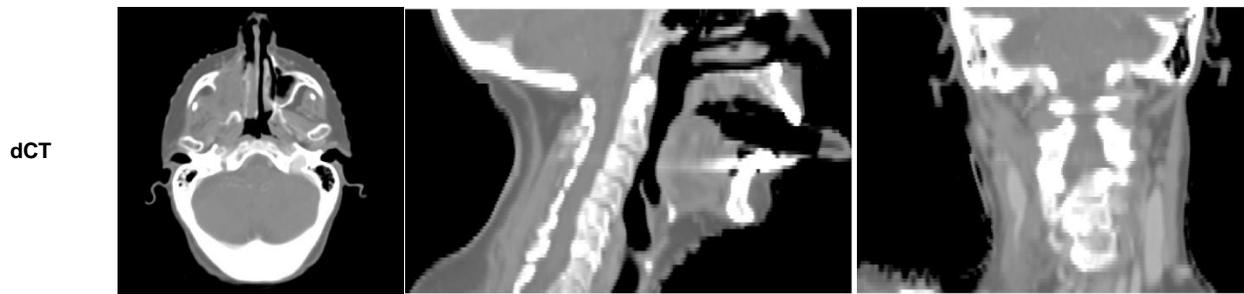

**dCT**

**Supplementary Figure 2.** Visualization of one randomly selected patient's CBCT, sCT and dCT images (H&N2). The top row shows CBCT images; the middle row shows sCT images generated from CBCT by the source model; and the bottom row shows the dCT images used as the ground truth. The left column shows images from the axial view; the middle column shows images from the sagittal view; and the right column shows images from the coronal view. The display window is from -400 HU to 400 HU.

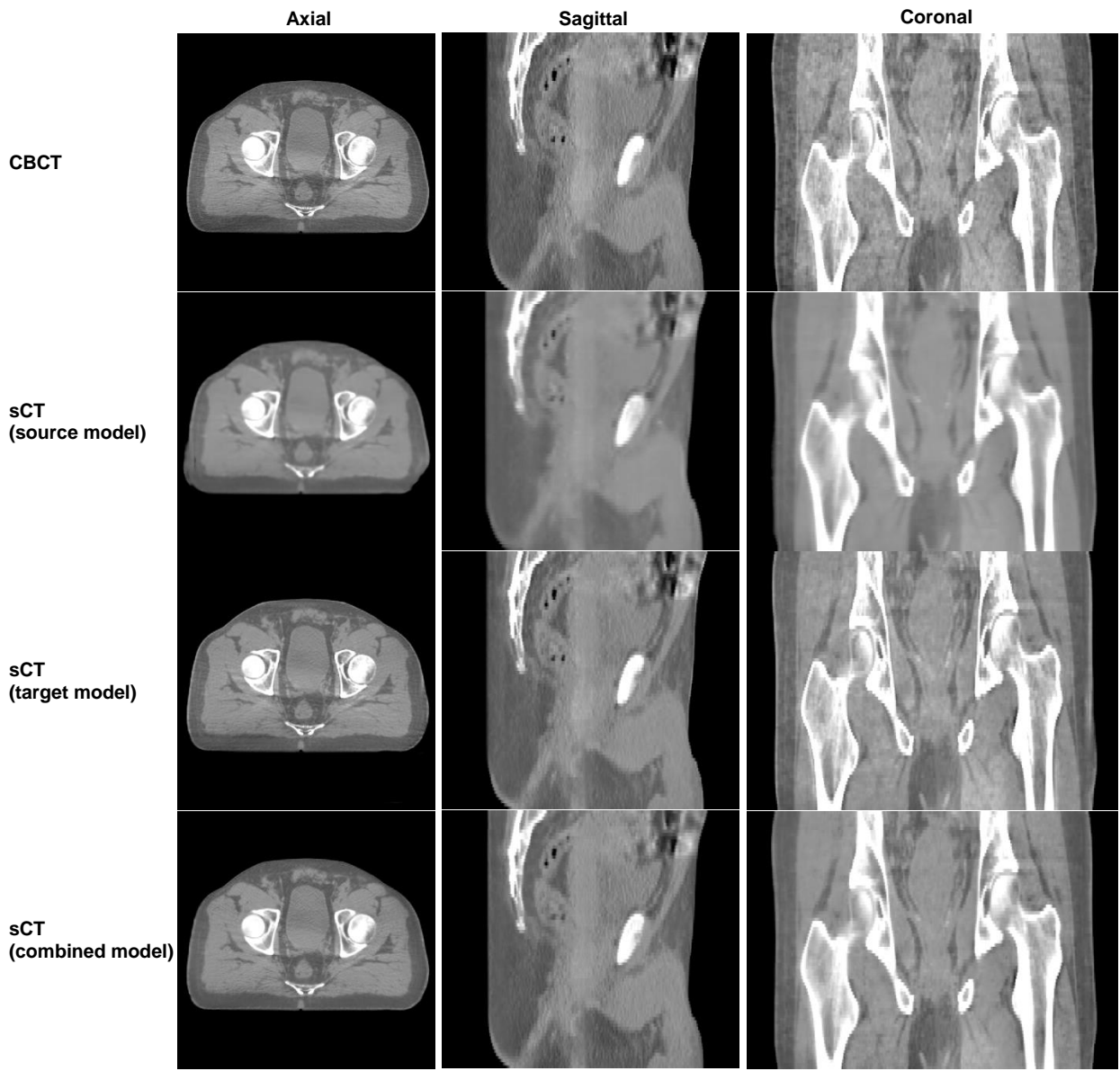

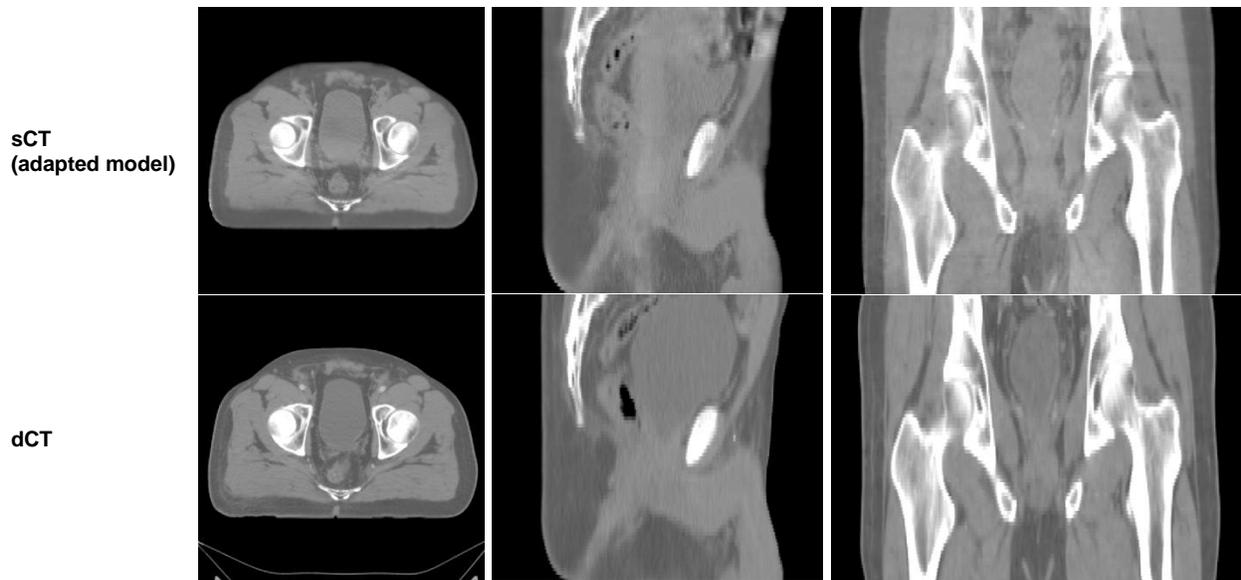

**Supplementary Figure 3.** Visualization of one randomly selected patient's CBCT, sCT and dCT images (Prostate1). The top row shows CBCT images; the second, third, fourth, and fifth rows show sCT images generated from CBCT by the source model, target model, combined model, and adapted model, respectively; and the bottom row shows the dCT images used as the ground truth. The left column shows images from the axial view; the middle column shows images from the sagittal view; and the right column shows images from the coronal view. The display window is from -400 HU to 400 HU.

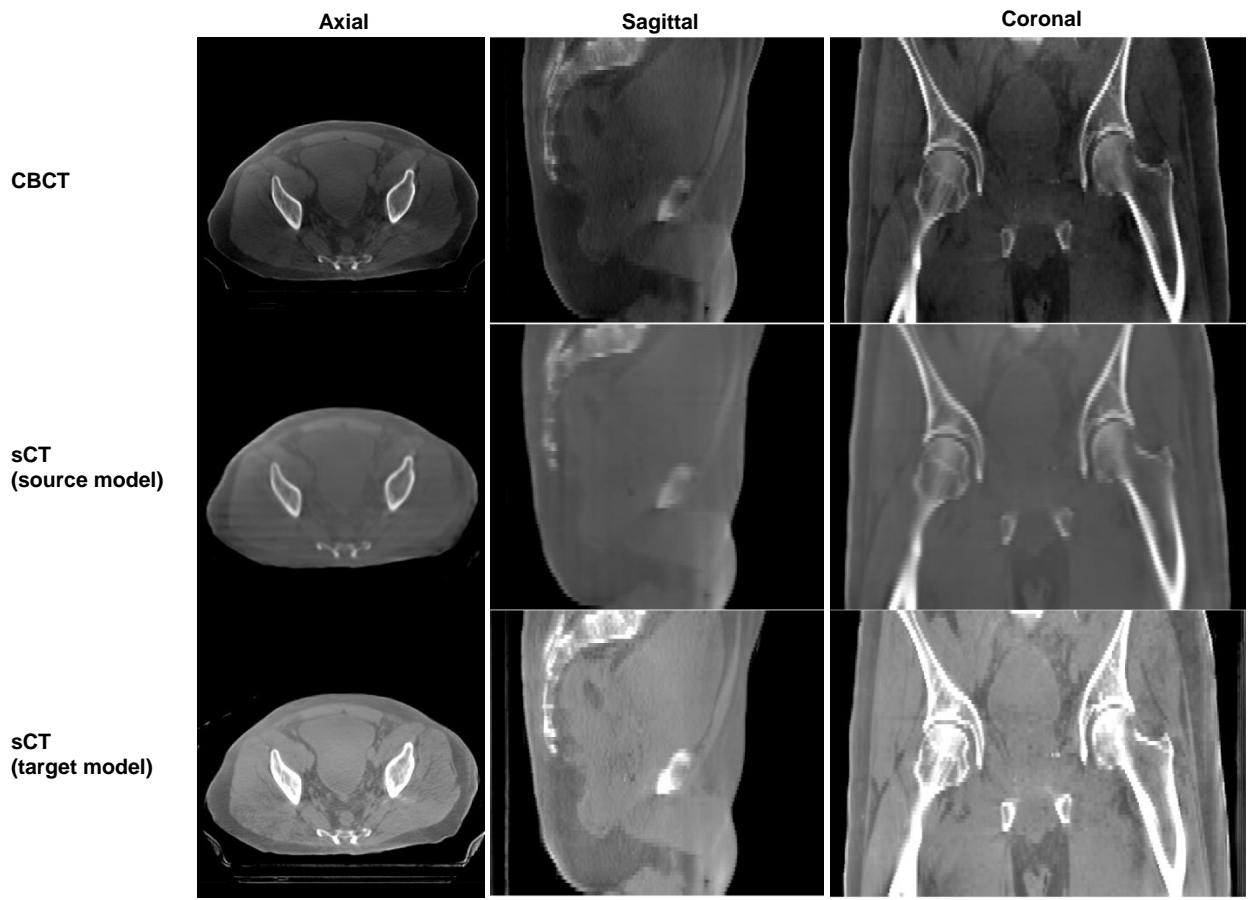

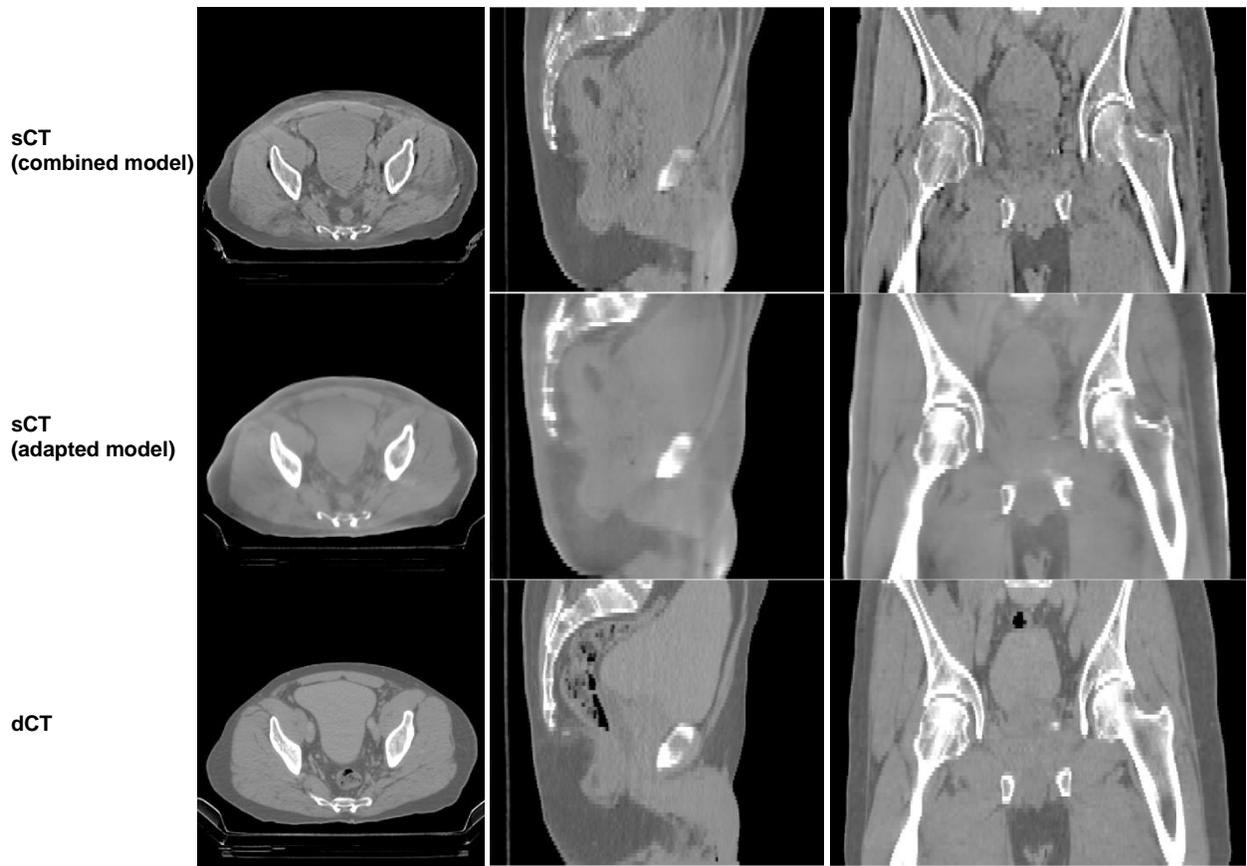

**Supplementary Figure 4.** Visualization of one randomly selected patient's CBCT, sCT and dCT images (Prostate2). The top row shows CBCT images; the second, third, fourth, and fifth rows show sCT images generated from CBCT by the source model, target model, combined model, and adapted model, respectively; and the bottom row shows the dCT images used as the ground truth. The left column shows images from the axial view; the middle column shows images from the sagittal view; and the right column shows images from the coronal view. The display window is from -400 HU to 400 HU.

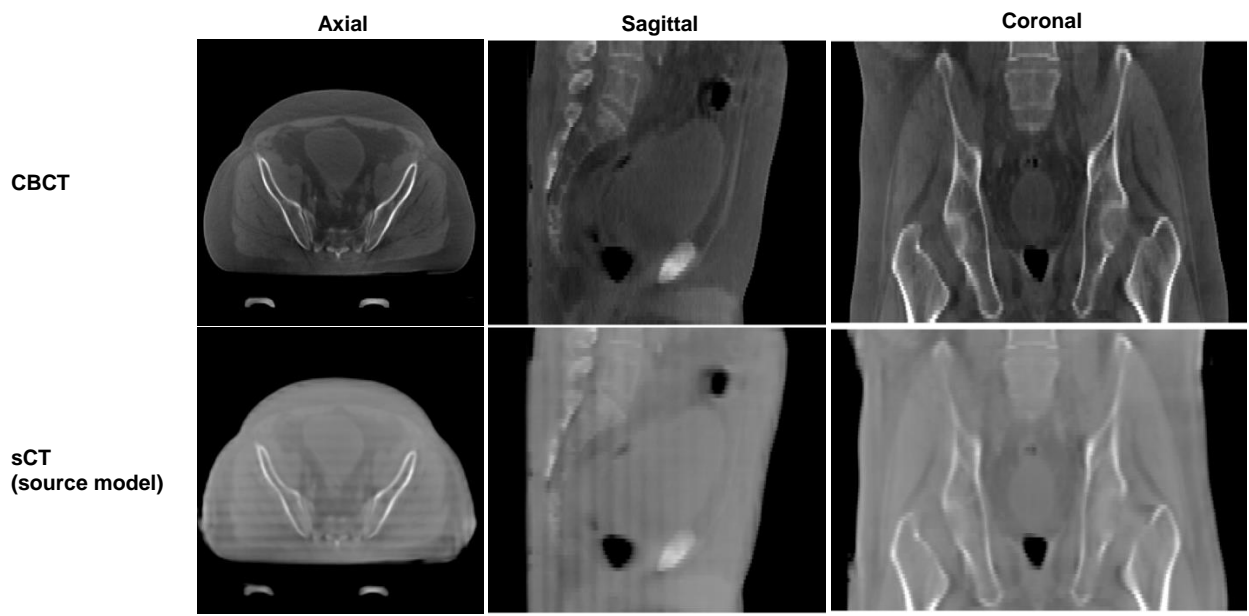

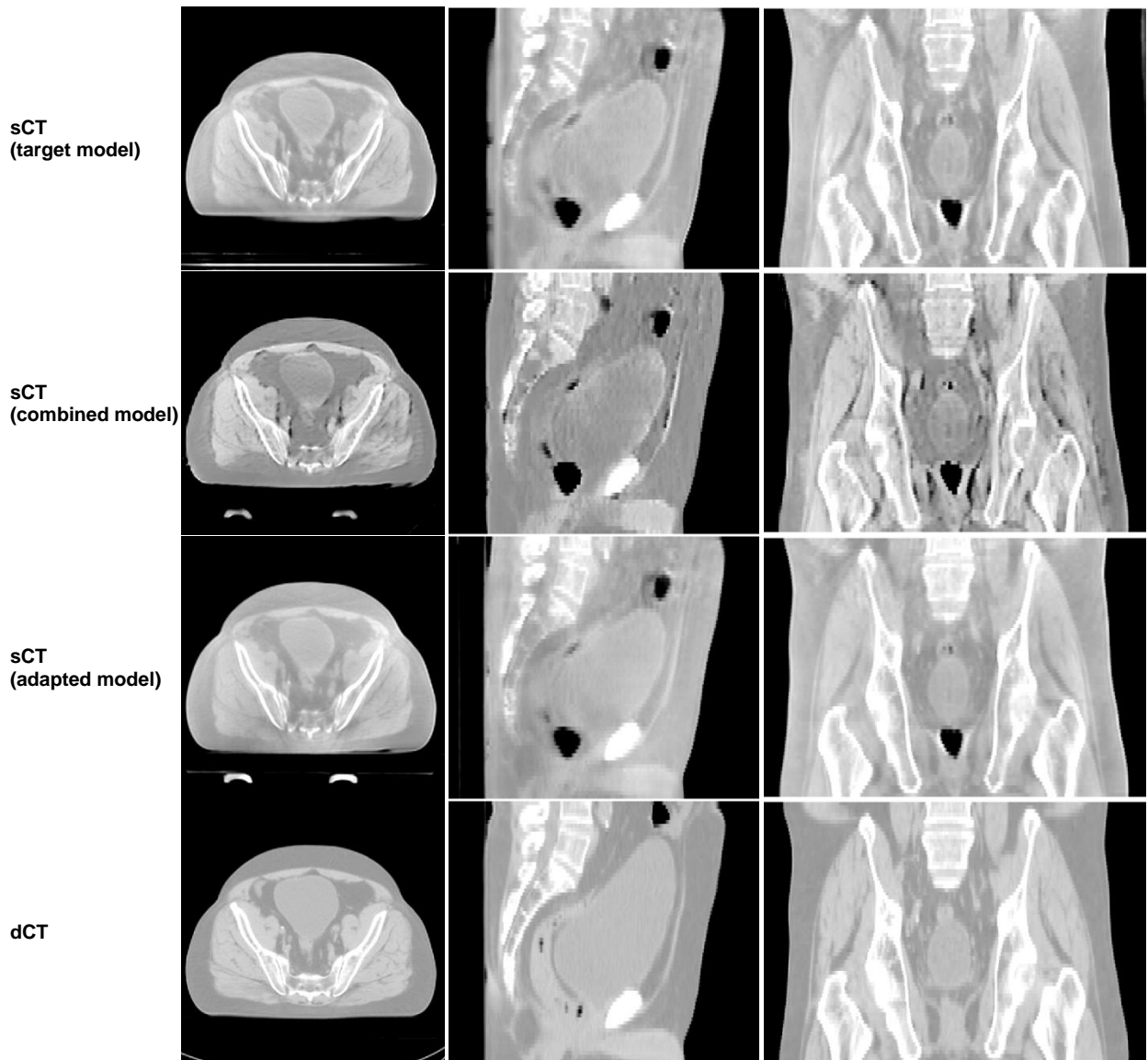

**Supplementary Figure 5.** Visualization of one randomly selected patient's CBCT, sCT and dCT images (Prostate3). The top row shows CBCT images; the second, third, fourth, and fifth rows show sCT images generated from CBCT by the source model, target model, combined model, and adapted model, respectively; and the bottom row shows the dCT images used as the ground truth. The left column shows images from the axial view; the middle column shows images from the sagittal view; and the right column shows images from the coronal view. The display window is from -600 HU to 200 HU.

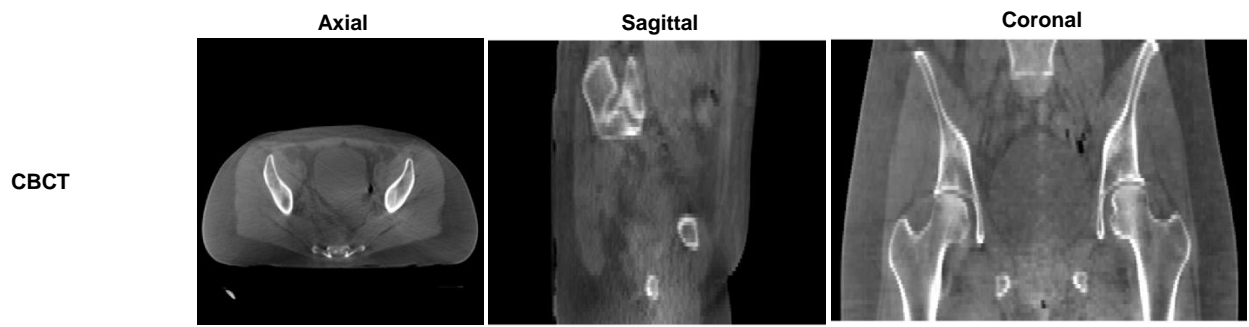

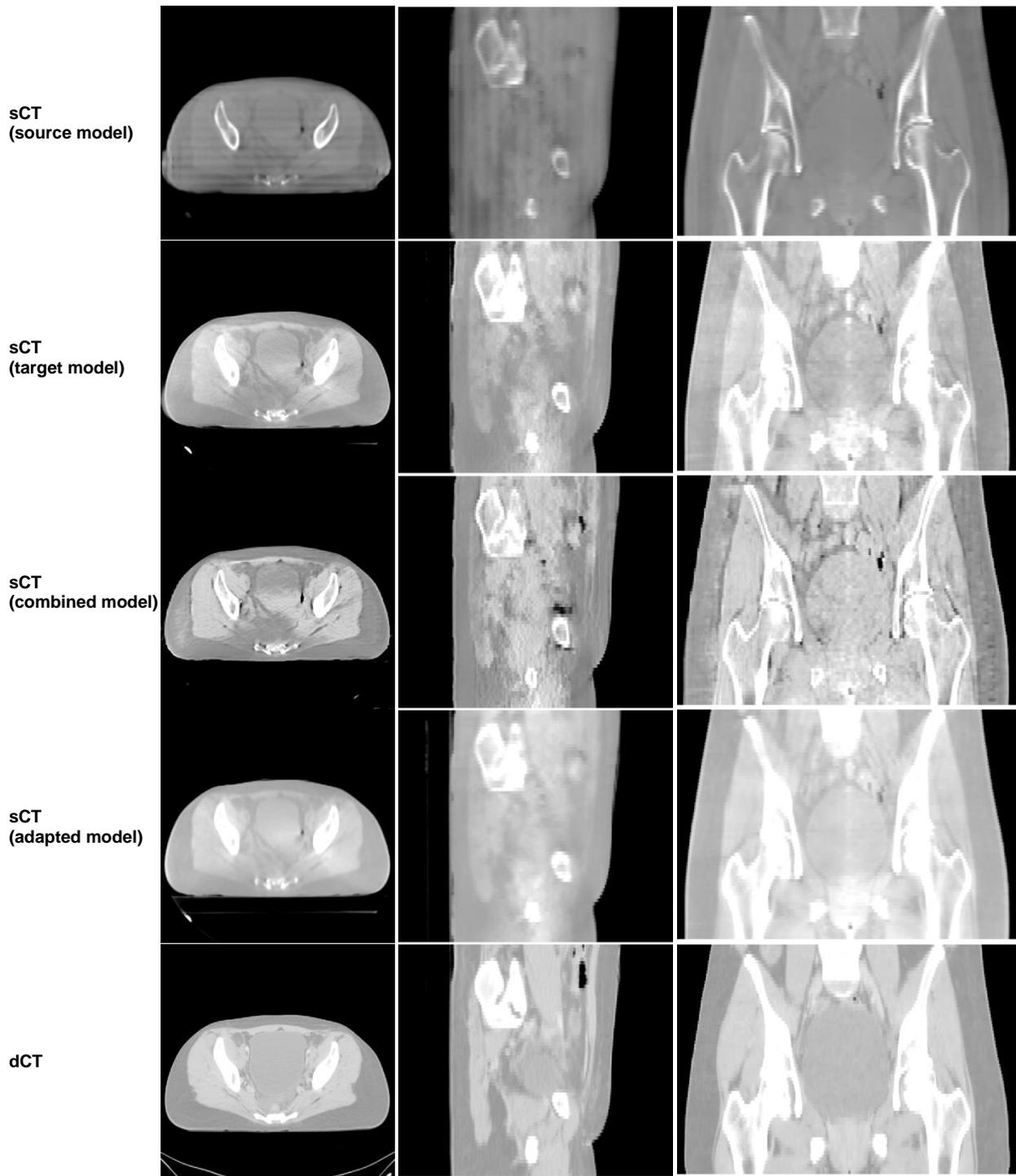

**Supplementary Figure 6.** Visualization of one randomly selected patient's CBCT, sCT and dCT images (Cervix). The top row shows CBCT images; the second, third, fourth, and fifth rows show sCT images generated from CBCT by the source model, target model, combined model, and adapted model, respectively; and the bottom row shows the dCT images used as the ground truth. The left column shows images from the axial view; the middle column shows images from the sagittal view; and the right column shows images from the coronal view. The display window is from -600 HU to 200 HU.

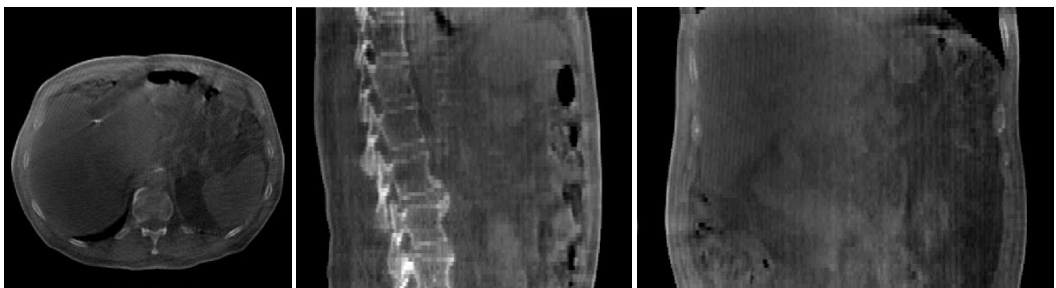
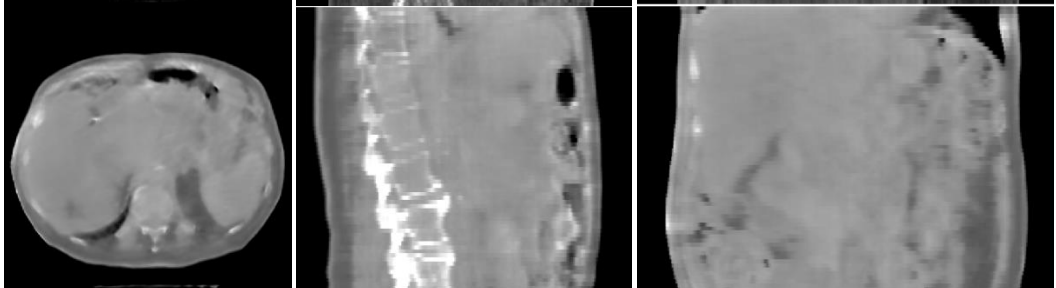
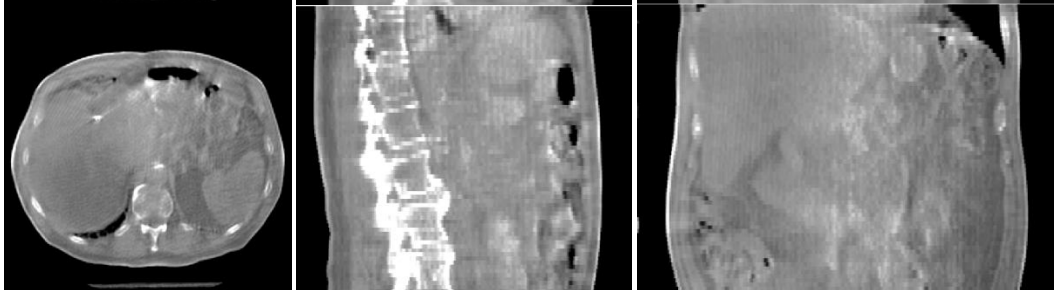
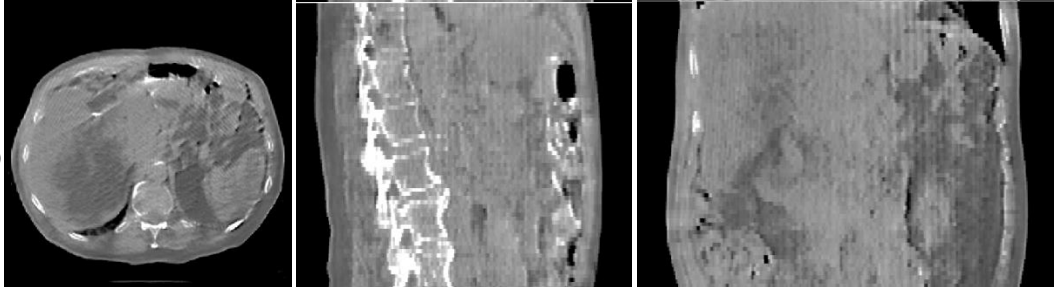
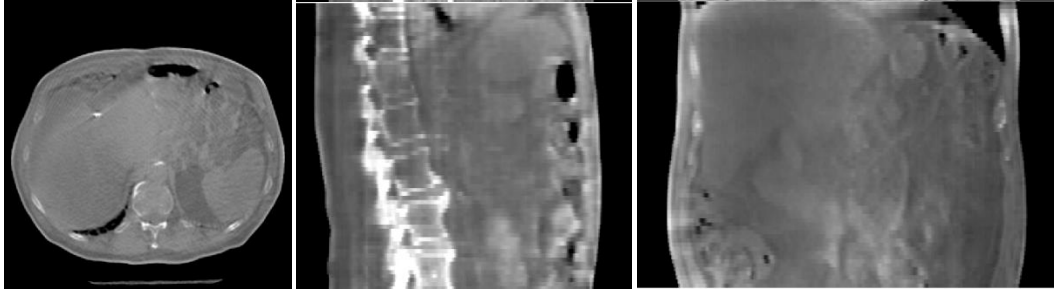

**CBCT**

**sCT (source model)**

**sCT (target model)**

**sCT (combined model)**

**sCT (adapted model)**

**dCT**

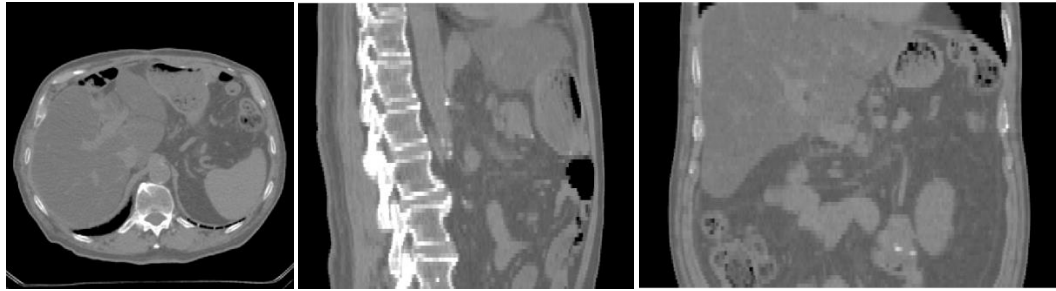

**Supplementary Figure 7.** Visualization of one randomly selected patient's CBCT, sCT and dCT images (Pancreas). The top row shows CBCT images; the second, third, fourth, and fifth rows show sCT images generated from CBCT by the source model, target model, combined model, and adapted model, respectively; and the bottom row shows the dCT images used as the ground truth. The left column shows images from the axial view; the middle column shows images from the sagittal view; and the right column shows images from the coronal view. The display window is from -400 HU to 400 HU.